\documentclass[conference]{IEEEtran}
\usepackage{latexsym}
\usepackage{graphicx}
\usepackage{mathptmx}
\usepackage{amsmath}
\usepackage{amsfonts}
\usepackage{amssymb}
\usepackage{amsbsy}
\usepackage{amsthm}
\usepackage{array}
\usepackage{multirow}
\usepackage{hhline}		
\usepackage{subfigure}
\usepackage{mathptmx}
\usepackage{mathtools}
\usepackage{todonotes}
\let\labelindent\relax
\usepackage{enumitem} 
\usepackage[square,sort&compress,comma,numbers]{natbib}

\ifCLASSINFOpdf

\else

\fi

\begin{document}
%
\title{A Novel Approach to Solve K-Center Problems\\ with Geographical Placement \vspace*{-0.6cm}
}
\author{
\IEEEauthorblockN{Peter Hillmann, Tobias Uhlig, Gabi Dreo Rodosek, Oliver Rose}
\IEEEauthorblockA{Universit\"at der Bundeswehr M\"unchen\\
Neubiberg, 85577, GERMANY\\
Email: \{peter.hillmann, tobias.uhlig, gabi.dreo, oliver.rose\}@unibw.de \vspace*{0.2cm}
}
}
\maketitle

\section*{ABSTRACT}

The facility location problem is a well-known challenge in logistics that is proven to be NP-hard. In this paper we specifically simulate the geographical placement of facilities to provide adequate service to customers. Determining reasonable center locations is an important challenge for a management since it directly effects future service costs. Generally, the objective is to place the central nodes such that all customers have convenient access to them. We analyze the problem and compare different placement strategies and evaluate the number of required centers. We use several existing approaches and propose a new heuristic for the problem. For our experiments we consider various scenarios and employ simulation to evaluate the performance of the optimization algorithms. Our new optimization approach shows a significant improvement. The presented results are generally applicable to many domains, e.g., the placement of military bases, the planning of content delivery networks, or the placement of warehouses.

\IEEEpeerreviewmaketitle

\vspace*{-0.1cm}
\section{INTRODUCTION}\label{introduction}
\begin{figure}[hbtp]
{
\centering
\includegraphics[width=0.48\textwidth]{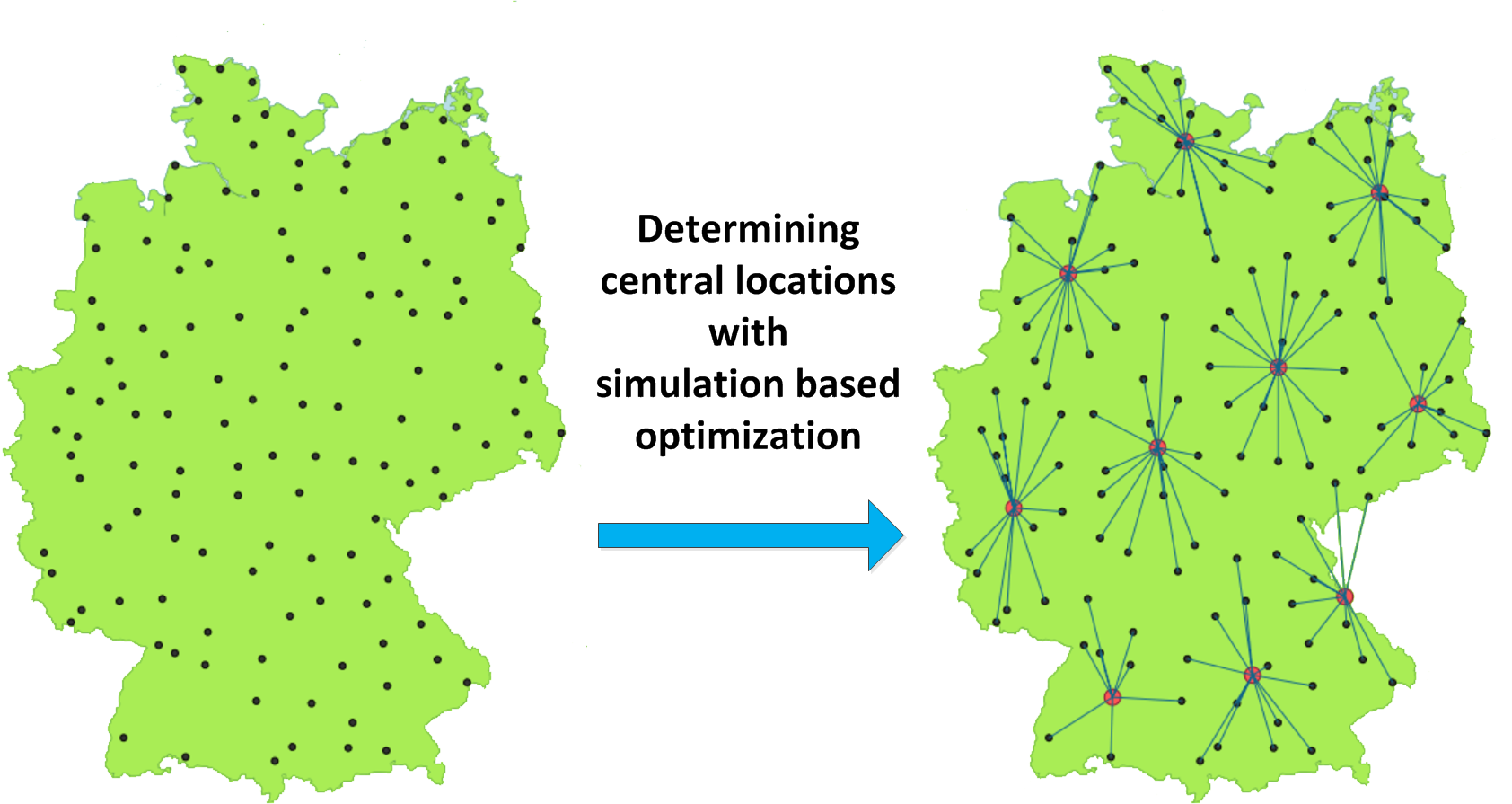}
\caption{Example scenario for placement of warehouses in Germany.}
\label{geographical}
}
\end{figure}

The typical geographical facility placement problem originates from the application area of logistic and transportation. Locations of central warehouses have to be determined with a short distance to  customers to provide adequate service and convenient access. An example is shown in Figure \ref{geographical}.
The left side shows the initial state with the geographical locations of customers. The scenario on the right side includes the optimized locations of ten warehouses with short distances to the customers. The edges represent the assignment of a customer to its nearest warehouse. 
The underlying theoretical problem is of importance in many different application areas \cite{Kanungo2002,Ganesh2010}, for example, the strategic placement of fire stations, hospitals, and military bases; as well as in content delivery networks for storage server allocation; and in telecommunication for improvements on the network infrastructure. Numerous further application areas with the same problem can profit from a good solution. 


Once a facility is set up at a specific location, it is very complex to change it. Facility location decisions are costly and have a strong long-term effect. Often only the set up costs and the capacity at a possible location of a facility is taken into account for the decision. The distance of a facility to its customers often receives less consideration, since existing placement strategies offer modest performance. The planning of the facility placement is very important and shall be well elaborated. Generally a facility should be placed close to the customers to reduce transportation time and cost. 
A well-planned facility improves the overall supply with service and goods. Therefor, the focus of this work is the geographical and infrastructural placement of facilities with a short distance to the customers. 
The facility location problem is based on the classic k-center and clustering problem.
In the uncapacitated version, it is assumed that a facility can provide service to as many customers as related to them \cite{Khuller00}. Nevertheless, the customers shall be more or less equally assigned to the facilities to obtain an equal load balancing. In the more complex version, the facilities have a predefined capacity to satisfy a limited number of customers \cite{Jain2001}. 


We identified three fundamentally different constraints for the placement of facilities in relation to the infrastructure. These are:
\begin{itemize} [itemsep=4pt] 
\item Free placement: The location can be determined completely free and without any restriction. 
\item Infrastructural placement: Existing infrastructure should be used for the placement. 
\item Node placement: Only a limited amount of given locations are viable to place a facility. 
\end{itemize}
Case one offers the largest degree of freedom, while the last case is the most restrictive approach. We analyze these constraints to illustrate the differences between them. 

In this article, we present a deterministic heuristic that finds optimized geographical locations of multiple, central locations. The approach is used for all three constraints. We compare the performance of our solution using benchmark scenarios as well as realistic scenarios.
We discuss the influence of management decisions, regarding the placement restrictions and the number of available facilities. To this end, we use simulation based optimization to analyze the necessary amount of facilities for an effective supply.

\section{SCENARIO AND REQUIREMENTS}\label{scenario}
The management of a large company intends to expand their business in another country and wants to distribute their products. 
The small supply chain in our example consists of three parts. Goods are produced in a factory overseas. From there, these products are transported to intermediate storages close to the distributor locations, where they are distributed to local stores for selling. The geographical location of the stores to be supplied and the precise infrastructure is known beforehand.
To implement this supply chain, multiple small warehouses need to be set up to provide a continuous supply of goods to the stores. Generally, the company aims for short transportation paths from their warehouses to the customers to reduce transportation costs and transportation times. These costs are balanced against the operational costs of multiple warehouses.
The placement of the warehouses and the assignment of stores to their closest warehouse have to be simulated and optimized automatically. For the initial planning phase, we consider only the uncapacitated geographical facility location problem. 
According to the scenario and various application areas, we need to answer the following important questions:
\begin{itemize} [itemsep=4pt] 
\item Where should new warehouses be placed geographically in order to obtain short and efficient transportation paths? 
\item Which consumer is assigned to which warehouse?
\item What is the necessary amount of warehouses to minimize the cost of operation and transportation?
\item How large is the performance difference between the constraint-free and the node placement scenario? 
\end{itemize}


\section{PROBLEM DESCRIPTION}
The scenario described in Section \ref{scenario} is based on the k-center or k-median problem, depending on the objective function. The uncapacitated facility location problem was first described in \cite{Balinski1965}. For a given amount of $V$ locations, a predefined amount of $K$ central locations have to be found. The k-center problem considers the minimization of the maximum distance between a location and its nearest center \cite{Chaudhuri1998}. In contrast, k-median problem uses the median as objective criterion \cite{Jain2001}. Both problems are NP-hard \cite{Gonzalez85}. The problem can be specified using a strongly connected graph topology \mbox{$G(V,E)$} with vertices ($v_{ i } \in V{}\ with{}\ i :=\{1,..,n\}$) and edges ($e_{ p } \in E{}\ with{}\ p :=\{1,..,m\}$).
The k-center problem is defined on a complete, undirected graph. The objective function $d$ defines the fitness value d($v_i$,$v_j$) for an edge e($v_{i}$,$v_{j}$) between two vertices ($v_{i}$,$v_{j}$), satisfying the triangle inequality. The objective function $d$ selects the best edge from a vertex $v_{i}$ to one of the calculated locations of a center node ($k_{u} \in K{}\ with{}\ u :=\{1,..,l\}\ and{}\ l \leq n $). A center node $k_{u}$ can be a vertice, dependent on the placement constraint from Section \ref{introduction}. The amount of all fitness values $d$ from every vertice $v_{i}$ is defined as $D$.

We intend to place the amount $K$ of center nodes to minimize the maximum $d$($v_{i}$, $k_{u}$) from a vertex $v_{i}$ to its \mbox{closest $k_{u}$.} This objective criterion belongs to the typical k-center problem according to Equation \ref{form:max}, the objective criterion of the similar k-median problem correspond to Equation \ref{form:med}:

\vspace*{-0.3cm}
\begin{equation}
D_{center}(K) = \underset{ } { min }\text{ } \underset{v_i = 1, ..., n }{ max } \text{ } \underset{ k_u \in K } {min}\text{ } d(v_i, k_u)
\label{form:max}
\end{equation}
\vspace*{-0.1cm}
\begin{equation}
D_{median}(K) = \underset{  } { min }\text{ }  \sum_{ v_i = 1 }^n { } \text{ } \underset{ k_u \in K } {min}   \text{ } d(v_i, k_u)
\label{form:med}
\end{equation}
\vspace*{-0.3cm}


Regarding our case, we set up a predefined amount $K$ of warehouses and minimize the maximum distance $D$ from the stores $V$ to its nearest warehouse $k$. The objective function $d$ defines the distance between a warehouse $k_{u}$ and a store $v_i$.



\section{RELATED WORK}\label{relatedwork}
Over the past years, many solutions for the facility location problem have been proposed. We focus on the most recent approaches and heuristics for comparison.
Most approaches analyzed the problem from the perspective of the k-center or clustering point of view with the objective to minimize the maximum distance \cite{Chaudhuri1998,Jain2000}. The papers of \citet{Rana09} and \citet{Arthur07} propose multiple heuristic approaches for the problem. Other important but more general work are from \citet{Potikas09}, \citet{Jamin01}, and \citet{Hochbaum85}. One of the current best solutions are from \citet{Resende2006}. The adapted greedy randomized adaptive search procedure (GRASP) meta heuristic combines a greedy initialization with a local search strategy. 
Besides these strategies there is a model of the node placement problem with mathematical relaxations \cite{
Pedroso2011}, which can be solved by a linear programming (LP) solver. We use the proposed approaches as benchmarks. Additionally, we use evolutionary algorithms like Simulated Annealing \cite{Alves1992} and a genetic algorithm \cite{Kratica2001}. The books of \citet{Klose2013}, \citet{Mayer2001}, and \citet{Fischer1997} present a comprehensive analysis of the logistical problem. But their focus is more on the economic aspects and the entire modeling process than on the adequate geographical placement of facilities.

The aforementioned work covers our requirements only partly or the proposed approaches show only a modest performance. They provide less information about the necessary amount of facilities to reach a specific objective.

\section{REFERENCE ALGORITHMS}
To compare our new approach with the existing strategies we present the most important reference algorithms. 
For our experiments, it is sufficient to consider the most restrictive case (node placement) and the most flexible approach (free placement). At the moment, we do not consider algorithms for infrastructural placement, since all algorithms can easily be adapted to generate appropriate solutions.
The algorithms in Table \ref{tab:algorithms} are suitable for certain placement constraints.
%
%
\begin{table} [htbp]
\begin{center} 
\caption{Overview of the presented algorithms and their usability. }
\label{tab:algorithms}
\begin{tabular}{|c|c|}
\hline
Node placement  & Node and Free placement \\
\hline
2-Approx &  k-Means  \\
Greedy & Evolutionary Algorithm  \\
GRASP &   \\
\hline
\end{tabular}
\end{center}
\end{table}
\vspace*{-0.7cm}


\subsection{2-Approx}
The 2-Approx choses a random vertice at the beginning, which becomes the location of the first center node. After that, the algorithm calculates for every placed center node the distance to all vertices. It chooses the vertice with the largest distance to their closest center node as the new location of the next center node. This iteration runs until the specified amount of center nodes is reached. 
This algorithm is 2-approximable.
Generally, $\beta$-approximable algorithm with factor $\beta$, guarantees a solution with cost $d$ where  $d \leq \beta  \cdot optimum$.
The algorithm guarantees that the maximum value of a distance from a vertice to its nearest center node is not larger than twice the maximum considering the optimal placement location of the center node \cite{Gonzalez85,Hochbaum85}. This bound is given without knowledge of the actual optimum. 




\subsection{k-Means}
The following group of clustering algorithms can be adapted for the free placement and node placement constraints. The main idea is to define k center nodes, one for each cluster. In this case a cluster is a group of vertices. Already placed nodes obviously influence the location of following placed nodes. In contrast to the iterative approach, these algorithms place all center nodes at the same time. This makes backtracking unnecessary. Placement changes of a center node have a direct effect to other center nodes.

Various approaches for clustering exist for a fixed number of clusters, they differ mainly with regard to the initial placement of center points. The \citet{MacQueen67} algorithm is one of the less complex k-Means algorithms. It relies on randomly selected locations of the vertices, which are used as the initial locations of the center nodes. Compared to the algorithm from \citet{Lloyd82}, also known as Voronoi relaxation, it starts with completely randomly placed center nodes in the area. Another typical initialization is used in the k-Means++ algorithm \cite{Arthur07}. Here, the location of the first center node is chosen randomly at a vertice location. The other center nodes are also placed randomly at vertice locations, however, the probability is skewed to favor certain locations. The selection probability is increased proportionally with their squared distances to already selected locations. 

After initialization, all vertices are assigned to their respective center nodes. For each group of vertices related to a center node an updated location is calculated. The new center node is the geometrical center of all vertices in a group. This process is repeated until center node locations do no longer change.
During every iteration step, the vertices are reassigned to the nearest center node. 
The algorithms are run multiple times due to random initialization. To adapt these strategies for the three constraints, the locations of the center nodes are mapped to the nearest possible location either in every step or at the end of the optimization.

\subsection{Greedy and GRASP}
The Greedy strategy \citet{Jamin01} initially places the center nodes at the location of predefined vertices iteratively. During each iteration it tries all possible placement combinations for the next center node and ultimately selects the location that provides the biggest benefit with respect to the optimization criterion. The Greedy strategy repeats this process until all center nodes are placed. 
To improve the quality further they include backtracking, to test whether already placed center nodes can be placed in a better way or may be removed completely.

The advanced approach of GRASP tries several starting locations as initialization for a greedy local optimization \cite{Pitsoulis2009
}. A weighted greedy randomized strategy is used for the initialization process. These starting locations are subsequently iteratively improved using local search. It updates the location of randomly chosen center nodes in a greedy manner.


\subsection{Evolutionary Algorithm}
Finally we use several evolutionary algorithms to optimize the center node locations. The SEREIN framework \cite{serein} is used to implement these algorithms. We employ the standard implementation of a genetic algorithm (GA) provided by SEREIN and use a population with 25 individuals evolving over 80 generations.
In addition, a Particle Swarm Optimization (PSO) and Simulated Annealing (SA) approach is implemented as well.
The parameters for the algorithms were determined experimentally using meta-optimization.


\section{OUR APPROACH DRAGOON}
Based on the k-Means strategy, we developed a new algorithm Dragoon (Diversification Rectifies Advanced Greedy Overdetermined Optimization N-Dimensions).
Most established placement algorithms are very sensitive to the initial placement of centers. Furthermore, the first center placed usually serves a high amount of customers.  This is possible in the uncapacitated facility location problem, but it shows the serious influence of the first placement decision. Nevertheless, an even distribution would be desirable. After the initialization, the vertices are assigned to the nearest center node location. In an iterative optimization these locations are improved. Most existing approaches try to find optimized locations only with respect to these single groups. The influence to other groups and the overall system is lost, which leads to suboptimal solutions. With the knowledge of these problem properties and weaknesses of current solutions, we designed our own algorithm. The algorithm should avoid as much as possible random decisions to prevent multiple runs and to reach stable results.



To reduce the sensitivity to initialization, we designed a new initialization process.
In the preliminary stage of the initialization phase, an orientation node is placed at the optimal position comparable to the one center node case. To avoid complete search, this can be simplified by calculating the average value of the coordinates. Afterwards, the specified amount of center nodes is placed using the 2-Approx strategy. Thereby, we obtain a very specific solution of the 2-Approx placement strategy. This guarantees the 2-approximable quality of the result.
After the initialization, the algorithm starts with the iterative refinement. These newly designed optimization steps are adaptable to different placement constraints.

The following description explains the general approach, which is adaptable to all three placement constraints. In every iteration step, the vertices are (re)assigned to their nearest center nodes. Afterwards, an updated location is calculated for every cluster of vertices related to a center node. This is done with respect to the entire scenario. The algorithm tests all possible locations around the current center restricted by the current cluster. The new location is chosen after the best improvement. This is done with respect to the specified optimization criterion. In our case, it is the maximum distance. If this value is unchanged, the algorithm will use another additional criterion. To choose between two solutions and to identify an improvement, we use an average or mean criterion. In each iteration, every center node is allowed to shift its location only once. This leads to a stepwise improvement and avoids a too fast stagnation in a local optimum. The order of the cluster selection has mostly no influence on the final result. This is due to the global view. 
For our simulations, the clusters are chosen with respect to their worst performance first. This iterative optimization is repeated until all center node locations do not change any more. Usually, only a few iteration steps are necessary until the algorithm terminates due to the described initialization. The algorithm accepts only improved locations in every iteration step. Therefore, the 2-approximable condition holds and it will always terminate. 

\begin{figure}[htb]
{
\centering
\includegraphics[width=0.48\textwidth]{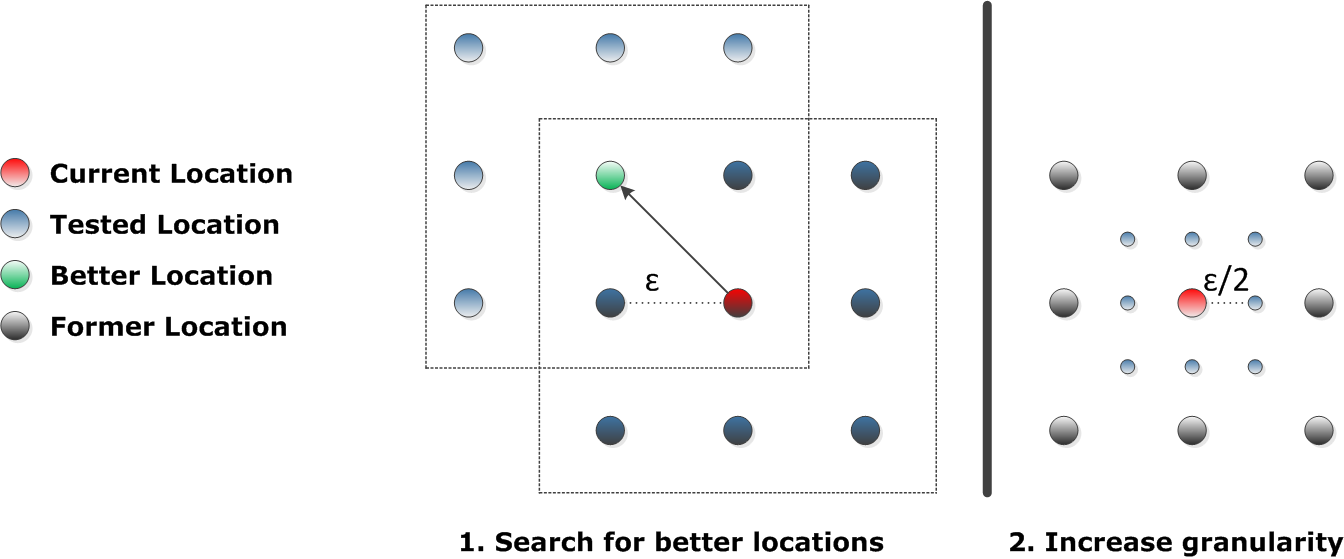}
\caption{Iterative optimization stage of the algorithm Dragoon by free placement constraint.}
\label{Dragoon}
}
\end{figure}

For the free placement constraint, our algorithm tests all points on a grid with a defined distance ($\epsilon$). If one of the tested locations results in a better performance for the overall scenario, it will be accepted. This location is used for the next iteration step. If no location leads to an improvement, we successively increase the granularity of the grid ($\epsilon_{new} := \epsilon_{old} / 2$). This process is repeated until the grid distance $\epsilon$ is smaller than the maximal accepted deviation. It is necessary to define a limit for the maximal deviation to terminate the optimization process. The processing steps of the iterative optimization are shown in Figure \ref{Dragoon}. The left side illustrates the movement to an improved spot. The right side shows the increased granularity of the grid by bisection.

For the node placement constraint, the algorithm simply tests all locations of grouped vertices for a center. To identify an improved location, the algorithm evaluates the overall scenario. All actual center locations are used in every evaluation including reassignment except of the observed center node. The possibilities of better center node location are limited to the group in each iteration step. To improve the performance, the tested capabilities can be further restricted to locations with a defined distance to the current center. The algorithm optimizes the center locations iteratively until no changes occur. According to the divide and conquer principle, the one center problem is solved optimal for each group of vertices with respect to the overall scenario. This optimization is calculated in polynomial runtime.


The algorithm can also be adapted to upgrade an existing scenario with partly fixed centers from the beginning or other constraints. A typical application area for this algorithm is the clustering of data.


\section{SIMULATION AND ASSESSMENT}\label{simulation}
The evaluation of the algorithms is based on experiments using a prototypic implementation in Java 
We use classic geo-coordinates in the 2-dimensional space and the euclidean distance as metric. 
We use more than 10 scenarios with equally weighted vertices and edges. The test set consists of a randomly generated and realistic scenarios without hierarchical topologies or other particular conditions. For the evaluation, the calculated distances of the different scenarios are normalized for comparability. 
As fitness function we calculated the distance parameter: maximum, 95\% quantile, median and average. To guarantee statistical significance we repeated simulations using multiple scenarios with an amount of vertices from 600 to 1200. 
The achieved results are accumulated for each algorithm.

Initially, a Monte Carlo approach for node placement serves as a basic benchmark. It shows that the 2-Approx algorithm on average returns much better and stable results. 
So we use the 2-Approx as reference.
Based on the results $s$ of the 2-Approx we define a theoretical limit for the optimum:    
\begin{equation}
 s \leq 2 \cdot Optimum_{ real} \implies Optimum_{ theoretical} \geq \frac{ 1 }{ 2 } \cdot s
\end{equation}
The solutions of 2-Approx vary because of the random initialization. Nevertheless, the 2-approximable condition is valid every run.

\begin{figure}[hbtp]
\centering
   \includegraphics[width=0.41\textwidth]{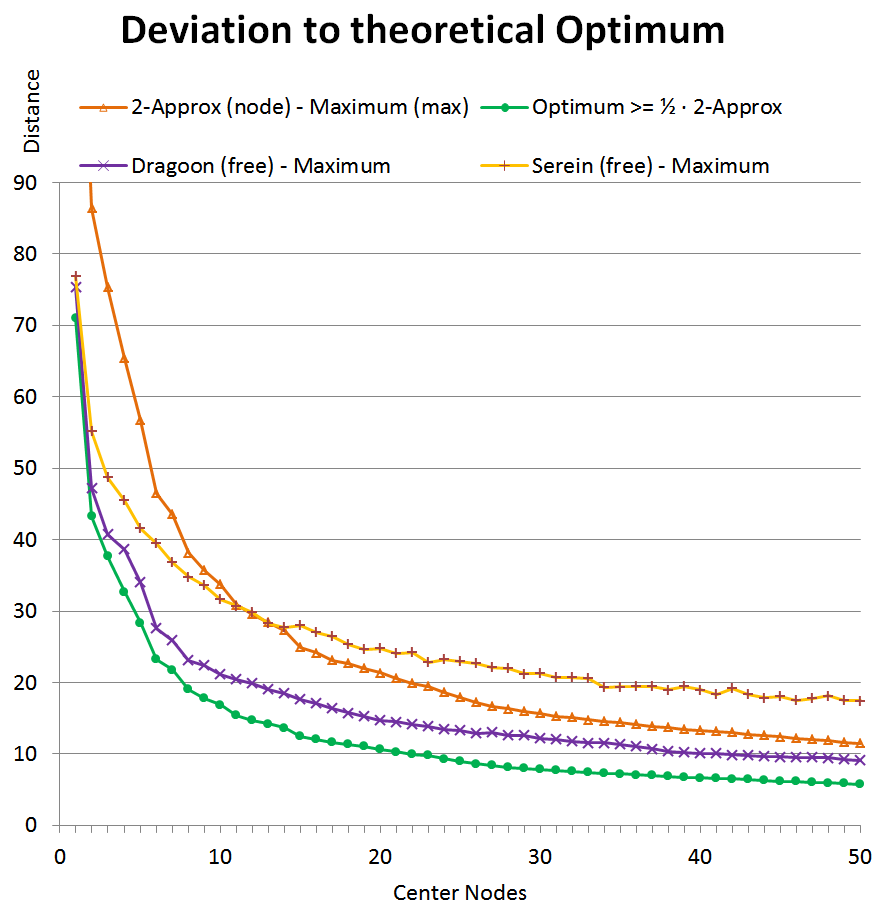}
   \caption{Deviation between 2-Approx and our best algorithms for maximum distance.}
   \label{DeviationToTheorecticalOptimum}
\end{figure}

Figure \ref{DeviationToTheorecticalOptimum} shows the results of the algorithm comparison. For small center node amounts, our improved Dragoon algorithm is close to the theoretical optimum. 
For larger center node amounts the algorithms stagnate with their performance nearly at the same level, referring that we are already very close to the actual optimum. Our approach performs significantly better than 2-Approx and is much faster than a brute force approach.

\begin{figure}[hbtp]
\centering
  \includegraphics[width=0.42\textwidth]{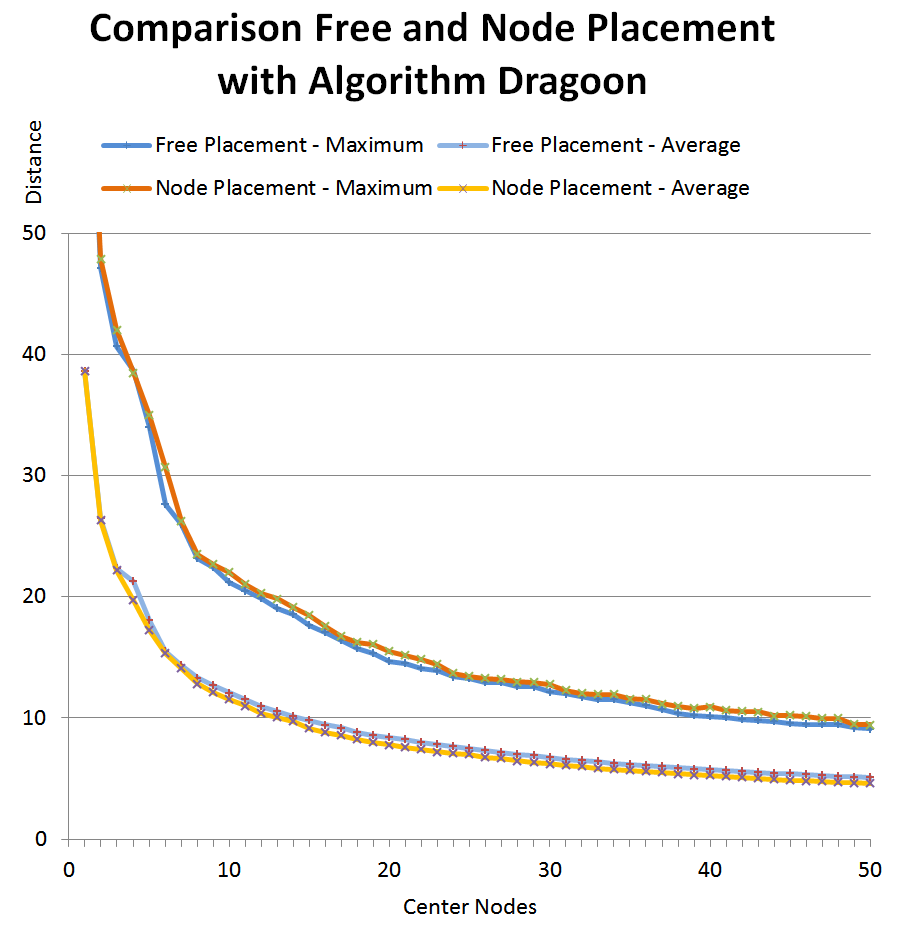}
  \caption{Deviation between free and node placement for maximum and average distance.\newline\newline}
  \label{ComparisonFreeAndNodePlacement}
\end{figure}

Figure \ref{ComparisonFreeAndNodePlacement} presents the difference between free placement and node placement for our improved algorithm Dragoon. The distance deviations between the different placement constraints are on average 4\% and in the worst case 11\%. We observe that the node placement approach needs on average 2 centers more to compensate the more flexible positioning of the free placement. In the worst case, 6 centers more are required. While additional center nodes have a positive effect, increasing the overall capacity and load balance, the added benefit decreases significantly for large amounts of center nodes. We observe a saturation effect for high ratios of center nodes in relation to vertices.

\vspace*{-0.57cm}
 
\begin{figure}[hbtp]
\centering
   \includegraphics[width=0.45\textwidth]{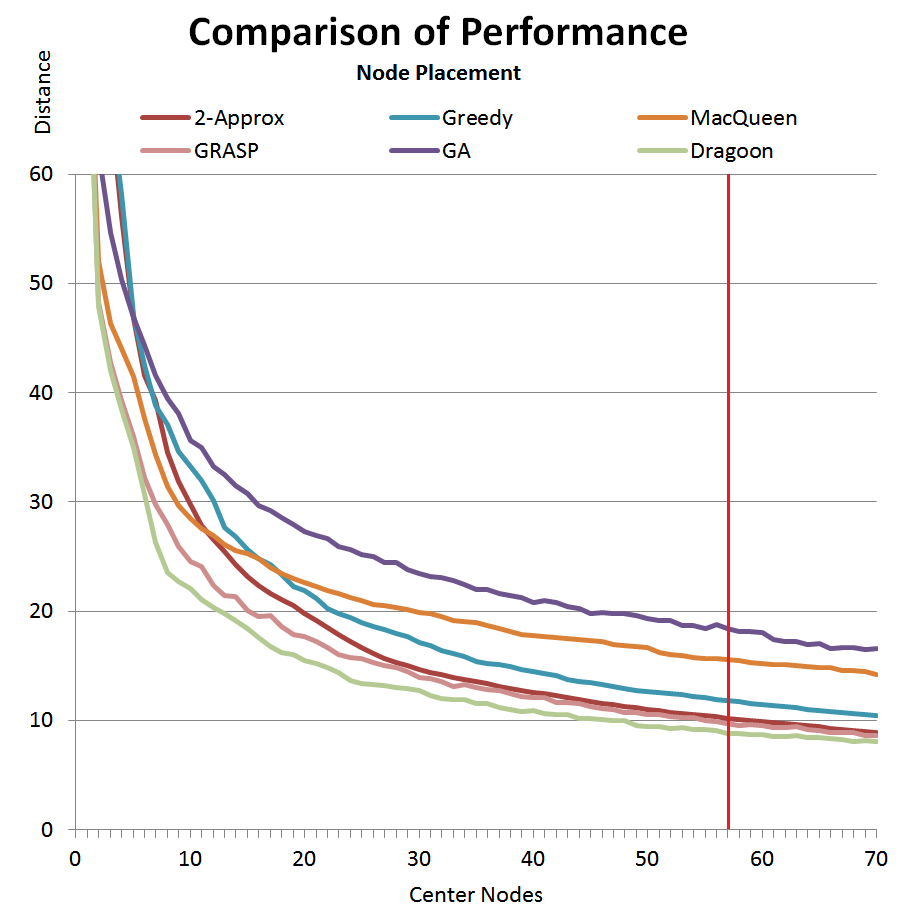}
   \caption{Performance overview for maximum distance. The red vertical line marks the amount of center nodes (57), after which the average performance improvement is less than 1\%.}
   \label{fig:PerformanceComparisonMax}
\end{figure}

Based on the maximum distance of a vertex to its nearest center node, Table \ref{tab:PerformanceComparisonMax} and Figure \ref{fig:PerformanceComparisonMax} show that it is sufficient to set up about 6\% of the vertices as center nodes. After we placed 58 center nodes in the normalized scenarios, the average improvement of maximum distance for an additional center node is less than 1\% with the 2-Approx or Dragoon algorithms.

\begin{table} [htbp]
\centering
\caption{Improvement of maximum distance in relation to number of center nodes (2-Approx). }
\label{tab:PerformanceComparisonMax}
\setlength{\tabcolsep}{2pt}
\begin{tabular}{|c|c|c|c|c|c|c|c|c|c|c|c|}
\hline
Center Nodes & 1 & 2 & 5 & 10 & 20 & 30 & 40 & 50 & 60 & 70 & 80\\ \hline
Max Distance & 112.6 & 76.3 & 46.8 & 29.8 & 19.8 & 14.7 & 12.6 & 11.0 & 9.9 & 8.9 & 8.1 \\ \hline
Improvement in \% & - & 32.2 & 38.7 & 36.4 & 33.5 & 25.9 & 14.3 & 12.5 & 10.4 & 9.6 & 8.8 \\ \hline
\end{tabular}
\end{table}

Figure \ref{fig:PerformanceComparisonMax} and Table \ref{tab:PerformanceOverview} present the general performance of the different algorithms. The performance of the SEREIN framework with an evolutionary algorithms is remarkable. SEREIN is not customized for this problem but achieves good solutions in comparison to other algorithms specially developed for this task. 
The performance of the algorithms MacQueen, Lloyd and k-Means++ are nearly the same, so we took MacQueen, the best of the three. The complexity of the algorithms is considerably different, but all optimizations finished after a couple of minutes in every used scenario except LP. It took a much longer time, especially for large scenarios. 

\begin{table} [htbp]
\centering
\caption{Performance overview with objective maximum distance in relation to number of center nodes.\newline}
\label{tab:PerformanceOverview}
\setlength{\tabcolsep}{5pt}
\renewcommand{\arraystretch}{1}
\begin{tabular}{|c|c|c|c|c|c|c|c|c|c|}
\hline
\rotatebox{90}{Center Nodes} & \rotatebox{90}{Monte Carlo (node)} & \rotatebox{90}{2-Approx (node) \text{ }} & \rotatebox{90}{Greedy (node)} & \rotatebox{90}{MacQueen (node)} & \rotatebox{90}{MacQueen (free)} & \rotatebox{90}{GRASP (node)} & \rotatebox{90}{GA (node)} & \rotatebox{90}{\textbf{Dragoon (node)}} & \rotatebox{90}{\textbf{Dragoon (free)}} \\ \hline
 \hhline{----------}
1      & 85.7            & 112.6           & 76.0         & 76.9        &  77.0      &  76.0   & 83.0 & \textbf{76.0}      & \textbf{75.3}        \\ \hline
2      & 63.6            & 76.3             & 74.8         & 52.0        &  51.1      & 48.2    & 62.0 & \textbf{47.9}      & \textbf{47.1}        \\ \hline
5      & 48.4            & 46.8             & 47.1         & 41.4        &  40.1      & 36.0    & 46.9 & \textbf{35.0}      & \textbf{34.0}         \\ \hline
10     & 36.9           & 29.8             & 33.3         & 28.4        &  26.2      & 24.5    & 35.6 & \textbf{22.0}      & \textbf{21.2}         \\ \hline
20     & 27.8           & 19.8             & 21.9         & 22.6        &  20.5      & 17.7    & 27.3 & \textbf{15.5}      & \textbf{14.7}         \\ \hline
30     & 24.0           & 14.7             & 17.2         & 19.9        &  18.1      & 14.0    & 23.5 & \textbf{12.8}      & \textbf{12.1}         \\ \hline
40     & 21.3           & 12.6             & 14.4         & 17.8        &  15.9      & 12.1    & 20.8 & \textbf{10.9}      & \textbf{10.1}         \\ \hline
50     & 19.8           & 11.0             & 12.6         & 16.4        &  14.3      & 10.5    & 19.3 & \textbf{9.4}        & \textbf{9.1}         \\ \hline
60     & 14.7           & 9.9               &  11.5        & 15.2        &  13.3      &  9.6    & 18.0 & \textbf{8.7}        &\textbf{ 8.3}        \\ \hline
70     & 13.8           & 8.9               &  10.4        & 14.2        &  12.0      &  8.6    & 16.6 & \textbf{8.0}        & \textbf{7.7}         \\ \hline
80     & 12.9           & 8.1               &  9.6          & 13.5        &  11.5      &  8.0    & 15.3 & \textbf{7.5}        & \textbf{7.1}         \\ \hline

\end{tabular}
\vspace*{-0.3cm}
\end{table}

\begin{figure}[hbtp]
\centering
   \includegraphics[width=0.48\textwidth]{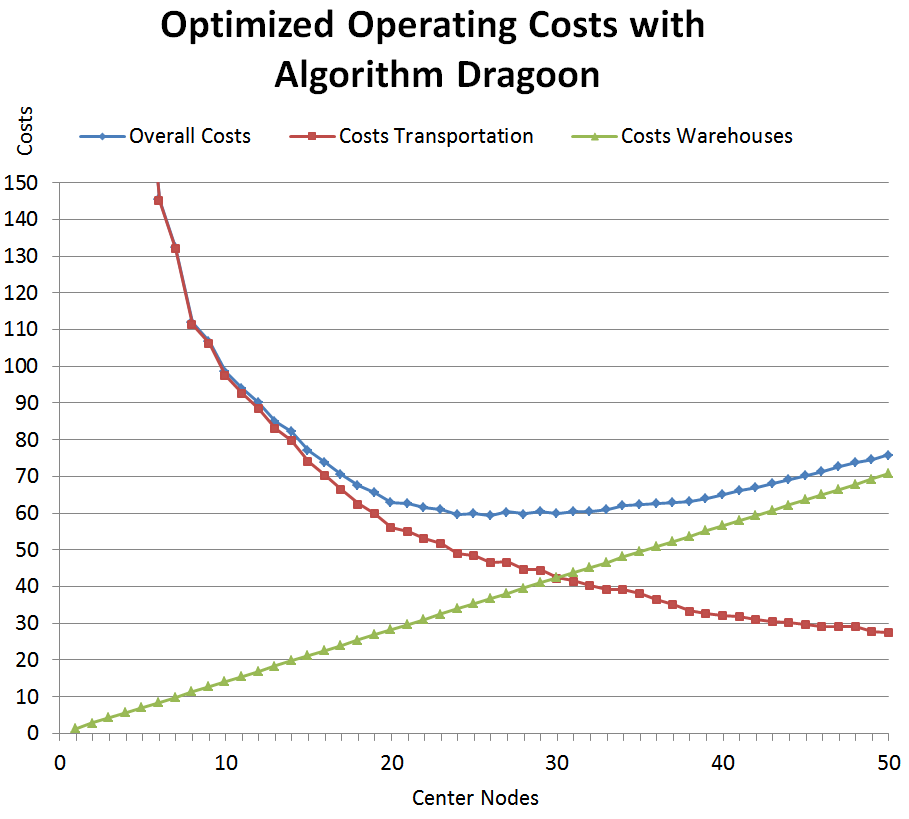}
   \caption{Optimized operating costs with our algorithm Dragoon. We reached the best relation of transportation costs and operating costs for warehouses by setting up 26 warehouses.}
   \label{costs}
   \vspace*{-0.3cm}
\end{figure}

In line with our initial intention, to set up warehouses for a specific scenario, the costs for transportation as well as operating and setup costs have to be respected. With an increasing amount of center nodes, the transportation distance and cost is reduced, whereas the set-up and operating cost is increased. To find the optimal balance between these aspects, we use simulation based optimization to calculate the optimal amount and location of center nodes. Figure \ref{costs} shows the operating costs for a specific scenario. This calculation has to be made for every scenario individually. For this example, we used abstracted cost function to show the objective of our simulation based optimization.

\section{CONCLUSION AND OUTLOOK}
In this paper, we propose the novel algorithm Dragoon to solve the k-center problem with geographical placement. Our strategy outperforms the other approaches, reaches very good results close to the global optimum in short time and is less sensitive to random initialization. 
We calculated the distance deviations between the different placement constraints (free vs node). 
A slight difference on average of 4\% is observed for the maximum distance. In the worst case, it can be up to 11\% distance difference between the most flexible case and most restrictive case.

To optimize the supply chain and the delivery time, we analyzed the amount of recommended center nodes for predefined scenarios. 
Our analyses show that even the best placement strategy reaches less than 1\% performance gain by adding an additional center node after an amount of a center node ratio of about 6\% is reached. 
In the future, we intend to further enhance the performance of the placement algorithms.
Furthermore, the inclusion of weights for customers and edges as well as different fitness functions will be considered.


\bibliographystyle{unsrtnat}
\bibliography{literature}
\end{document}